\documentclass[10pt,final,conference,letterpaper,twoside]{IEEEtran} 
\usepackage[hidelinks]{hyperref}
\usepackage{float}
\usepackage[caption=false,font=footnotesize]{subfig}
\usepackage{graphicx}
\usepackage{gensymb}
\usepackage{balance}
\usepackage{comment}
\makeatletter
\newcommand{\linebreakand}{%
  \end{@IEEEauthorhalign}
  \hfill\mbox{}\par
  \mbox{}\hfill\begin{@IEEEauthorhalign}
}
\makeatother
\begin{document}
\title{Assessment of Fetal and Maternal Well-Being During Pregnancy Using Passive Wearable Inertial Sensor}

\author{%
\IEEEauthorblockN{Eranda Somathilake}
\IEEEauthorblockA{\textit{Dept. of Mechanical Eng.} \\
\textit{University of Peradeniya}\\
Peradeniya, Sri Lanka\\
eranda.somathilake@eng.pdn.ac.lk}
\and
\IEEEauthorblockN{Upekha Delay}
\IEEEauthorblockA{\textit{Dept. Electrical \& Electronic Eng.} \\
\textit{University of Peradeniya}\\
Peradeniya, Sri Lanka\\
upekha.delay@eng.pdn.ac.lk}
\and
\IEEEauthorblockN{Janith Bandara Senanayaka}
\IEEEauthorblockA{\textit{Dept. Electrical \& Electronic Eng.} \\
\textit{University of Peradeniya}\\
Peradeniya, Sri Lanka\\
janith.b.senanayaka@eng.pdn.ac.lk}
\linebreakand
\IEEEauthorblockN{Samitha Gunarathne}
\IEEEauthorblockA{\textit{Dept. Electrical \& Electronic Eng.} \\
\textit{University of Peradeniya}\\
Peradeniya, Sri Lanka\\
samithalg@eng.pdn.ac.lk}
\and
\IEEEauthorblockN{Roshan Godaliyadda}
\IEEEauthorblockA{\textit{Dept. Electrical \& Electronic Eng.} \\
\textit{University of Peradeniya}\\
Peradeniya, Sri Lanka\\
roshangodd@ee.pdn.ac.lk}
\and
\IEEEauthorblockN{Parakrama Ekanayake}
\IEEEauthorblockA{\textit{Dept. Electrical \& Electronic Eng.} \\
\textit{University of Peradeniya}\\
Peradeniya, Sri Lanka\\
mpb.ekanayake@ee.pdn.ac.lk}
\linebreakand
\IEEEauthorblockN{Janaka Wijayakulasooriya}
\IEEEauthorblockA{\textit{Dept. Electrical \& Electronic Eng.} \\
\textit{University of Peradeniya}\\
Peradeniya, Sri Lanka\\
jan@ee.pdn.ac.lk}
\and
\IEEEauthorblockN{Chathura Rathnayake}
\IEEEauthorblockA{\textit{Dept. of Obstetrics and Gynaecology} \\
\textit{University of Peradeniya}\\
Sri Lanka \\
chathura67@hotmail.com}
}

\maketitle
\begin{abstract}
Assessing the health of both the fetus and mother is vital in preventing and identifying possible complications in pregnancy. This paper focuses on a device that can be used effectively by the mother herself with minimal supervision and provide a reasonable estimation of fetal and maternal health while being safe, comfortable, and easy to use. The device proposed uses a belt with a single accelerometer over the mother's uterus to record the required information. The device is expected to monitor both the mother and the fetus constantly over a long period and provide medical professionals with useful information, which they would otherwise overlook due to the low frequency that health monitoring is carried out at the present. The paper shows that simultaneous measurement of respiratory information of the mother and fetal movement is in fact possible even in the presence of mild interferences, which needs to be accounted for if the device is expected to be worn for extended times.
\end{abstract}

\begin{IEEEkeywords}
Fetal movement, breathing patterns, fetal health, accelerometers, deep learning, Weiner filtering, wavelet transform, fast Fourier transform
\end{IEEEkeywords}

\section{Introduction}
Regular monitoring throughout the pregnancy allows early detection of well-being problems that might arise and will aid their treatment, improving the chance for the birth of a healthy baby. The health and condition of both the mother and the baby is a clear indication of future complications or the well-being of the fetus \cite{health}. Fetal well-being can be monitored in  different ways \cite{fhl, fetal_sur, fetal_sur1}, each with its own advantages and disadvantages. Hence, the availability of multiple methods will provide the medical professionals with better tools, which can be used in the different specific situations. Also, the mood and health of the mother can have a dramatic impact on the development of the baby, both in long and short terms \cite{maternal_breathing}. This paper, therefore, proposes a device to assess the condition of the fetus as well as the mother, which is non-invasive, low cost, and simpler to use compared to most of the existing methods available for fetal condition monitoring.

Fetal condition monitoring can be done in multiple ways and the area of focus in this research is through fetal movement, which can be used as an indication of future complications \cite{movement}. Fetal movement can be identified as a primary indicator of fetal well-being. Reduction or absence of fetal movement is a strong indication of fetal compromise \cite{fetal_activity, fetal_activity2}. Fetal movement monitoring methods can be divided into two approaches: active or passive. Active methods directly observe the fetus using various imaging techniques while passive methods measure the fetal movements indirectly by measuring other responses as the movements of the fetus. CTG and Utrasound scanning are examples of active methods, whereas use of sensors such as accelerometers or acoustic sensors are examples of passive methods.

While most of the common active methods such as Ultrasound and CTG are used extensively, they do possess some drawbacks that can make them undesirable in certain situations. Although it is not proven clinically, the use of high frequency audio waves which penetrates the uterus may cause harm to the fetus \cite{non_invasive}. Also, the equipments used for both CTG scanning and ultrasound scanning are bulky and require trained professionals for operation and interpretation, thus making them impractical to be used in a daily basis over an extended period of time. Fetal state monitoring using these methods are only conducted in clinical settings and most of the time is done after the mother's admittance to the hospital.It would, however, be beneficial if fetal movements can be monitored domestically in an ambulatory setting, which will in turn enable the assessments to be done more frequently and is ideal in times of a pandemic, where the mothers can safely stay in their homes. In addition, it is difficult to obtain images of obese mothers, which makes these methods less viable to be used \cite{obesity} even though they are more likely to encounter complications \cite{obesity_complications}. The lack of experts and the expensive equipment required, in rural areas, can be a major hindrance \cite{rural} in monitoring fetal movement. Thus promoting the need for low cost and simpler methods. 

Simpler passive methods, therefore, can be used as an alternative or along with existing active methods to have a better assessment of fetal movement throughout the pregnancy. The proposed device detects fetal movement through an accelerometer. Obtaining the data through this device is more convenient, non-invasive, and takes minimal time for setup. This device is designed to be portable, ergonomic, and with modular electronics for easy repair. Furthermore, it was observed that the accelerometer placed on the abdomen have the ability to capture maternal respiratory motion data as well. It can, therefore, be used to detect respiratory patterns \cite{res_e1}, which can provide valuable information such as respiratory rate and maternal energy expenditure level to medical practitioners. Both respiratory rate \cite{res_imp} and maternal energy expenditure \cite{ee_pregnancy} are key indicators of maternal well-being during pregnancy. Hence, this device can be used as an overall condition monitoring device rather than solely focusing on fetal movement.

The data obtained using the sensors consist of information of many processes that happens in the body, namely fetal movements, maternal respiratory patterns, and disturbances such as coughing, laughing, or walking. In this study, the desired parameters were maternal respiratory motions and fetal movements. Therefore, analyses were conducted to identify fetal movements and to extract information on maternal respiratory data.
 

This device is designed to be used by lay users and the readings will not be taken in a controlled environment. Hence, the device should have the capability to identify and filter out unnecessary data. Therefore, in this study, it is demonstrated that the artefacts introduced into the accelerometric signal due to the walking during a session can be successfully removed using a Weiner filter. Moreover, a peak detecting algorithm was used to calculate the respiratory rate of the mothers, which can act as one of the main well-being indicators for maternal health \cite{res_e3}. Furthermore, analyses were conducted to classify the energy expenditure level of the mother's body, depending on the respiratory pattern. To accomplish that, feature extraction was done using the discrete Wavelet Transform and then classified using a Neural Network.

Fetal movement patterns can be complex in nature and hence machine learning techniques were used to identify them \cite{plos_one,fetal_lstm}. One approach was to use Convolutional Neural Networks (CNN), where a scalogram was generated by taking the wavelet transform of the signal and was then used to identify fetal movement. The other approach was to use a recurrent neural network, where Gated Recurrent Unit (GRU) cells were used so that the signal can be analysed over long term dependencies to detect fetal movement.

\section{Background}
Active methods such as Ultrasound scanning and CTG scanning observe the movement of the fetus and give a physical representation of it. They are therefore highly accurate and experts can assess fetal health fairly accurately \cite{health}. The negative aspects of these devices, as stated previously, are them being too complex to use, bulky, and might have harmful effects on both the mother and the baby. Furthermore, interpreting the data received from these methods requires technical skill as well as time, which can act as a hindrance to taking quick actions when necessary. Therefore, passive methods that observe the surface of the uterus was considered in this study. Also, active methods do not evaluate fetal movement constantly, but identification of changes in fetal movement patterns over an extended time period can help identify complications earlier \cite{fetal_activity2}. 

Maternal perception of movement identification and monitoring is a widely used and easy to implement method but it is highly subjective and inconsistent \cite{mat_per}. Hence, there is a need for a viable fetal movement monitoring method that eliminate these inaccuracies. One such method proposes a device that uses acoustic sensors and accelerometers attached to a belt worn around the abdomen \cite{acoustic}. Also, multiple accelerometer sensors were used in a different study \cite{recording_system, acc}. These multiple sensor approaches have yielded good results. However, as stated in \cite{acoustic}, acoustic sensors are too sensitive for this particular problem. The use of multiple accelerometers, although is better than one, did not seem to have a significant effect on movement detection and increasing the sensors had a diminishing impact on the results while it led to an increase in used equipment and complexity. Hence, in this study, a single sensor was used and more attention was paid to optimising the post-processing of the data. Most of the proposed devices mentioned above have performed well in an experimental environment. However, their capability to be adapted into practical situations where they are handled by users who are not technically trained were not considered in most cases. The proposed device in this study has been designed for practical implementation with ease of use and simplicity in mind.

Since inertial sensors measure the physical movement of the surface of the abdomen, in addition to capturing fetal movement data, they will simultaneously record other bodily functions of the mother as well. Hence, the device can observe multiple states of the mother. In this study, we have focused on monitoring the breathing patterns of the mother using the same inertial sensor. The use of accelerometers to measure breathing patterns is a viable solution and it has been implemented to classify between different breathing patterns \cite{res_e1}. A similar approach was considered in this study to evaluate the breathing patterns of pregnant mothers to evaluate their health.

\section{Device implemented}
\begin{figure}[!t]
\centering
\includegraphics[width=0.48\textwidth]{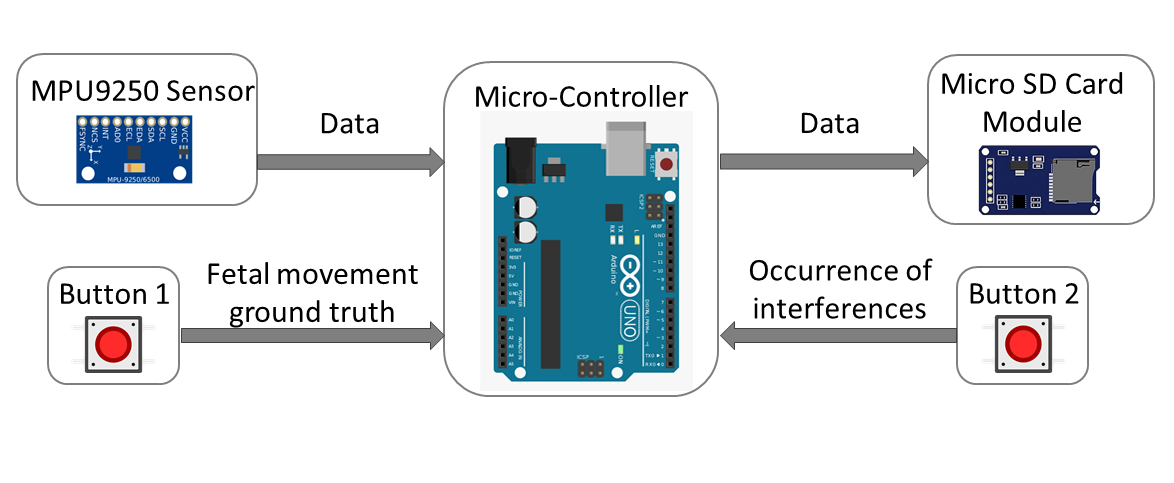}
\caption{Component configuration of the proposed device}
\label{fig:setup}
\end{figure}

The device implemented consists of an accelerometer and two buttons to get external inputs. One input was used by the doctor who observes the fetus using ultrasound scanning to be used as the ground-truth to train the neural network, while the other button is used to record other interference that might be detected by the accelerometer such as the mother laughing or coughing. The same device was used in separate sessions to record breathing patterns and this did not require any user input while recording. The configuration of the components of the device is given in \figurename~\ref{fig:setup}.

\begin{figure}[!t]
\centering
\includegraphics[width=0.48\textwidth]{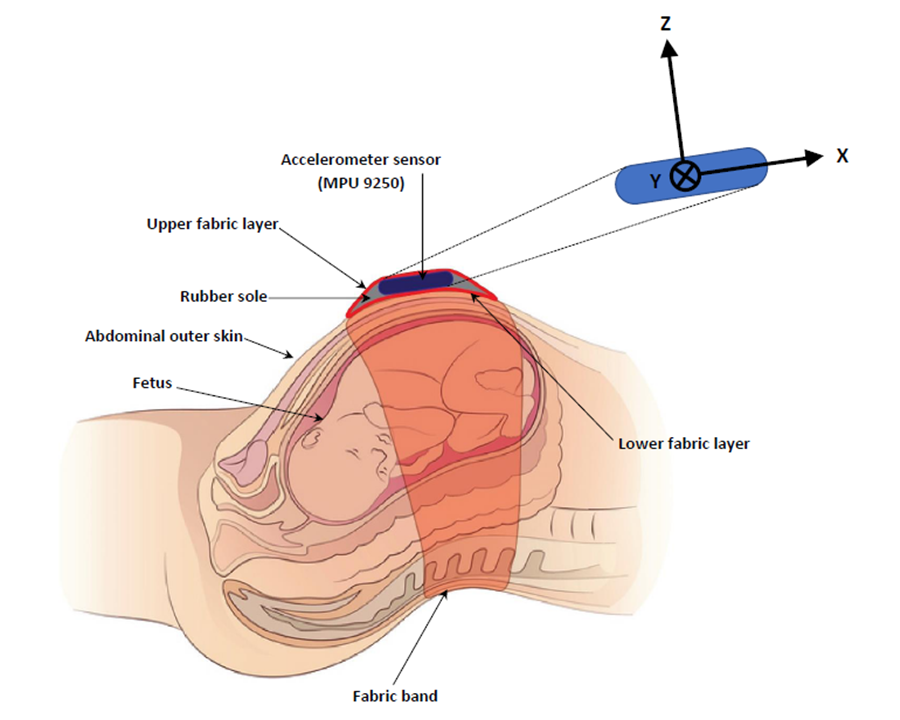}
\caption{Mother wearing the sensor}
\label{fig:worn}
\end{figure}

The components are controlled by an Arduino Uno micro-controller, as stated, the device consists of an accelerometer (MPU 9250) and two buttons. In addition, there is a micro SD card module to store data, a real-time clock (DS3231), and LEDs to indicate the current state of the device and button presses. The accelerometer was attached to a belt which can be wound around the mother's uterus as shown in figure \figurename~\ref{fig:worn}. In this preliminary study, the device was controlled by a PC to give better control over its operation.

The sensor used here, MPU 9250, consists of a tri-axial accelerometer, tri-axial gyroscope, and a compass. Only one axis of the tri-axial accelerometer was used in this analysis. The sensor communicates with the Arduino using the $I^2C$ interface up to a maximum sampling rate of \mbox{32 kHz}, but the readings obtained were at a sampling rate of around 100~Hz mainly due to the other processes running in the micro-controller. The specified temperature range for the operation of the sensor is from $-40\degree C$ to $85\degree C$, which accommodates the operating temperatures for this application.

\begin{figure}[!t]
\centering
\includegraphics[width=0.4\textwidth]{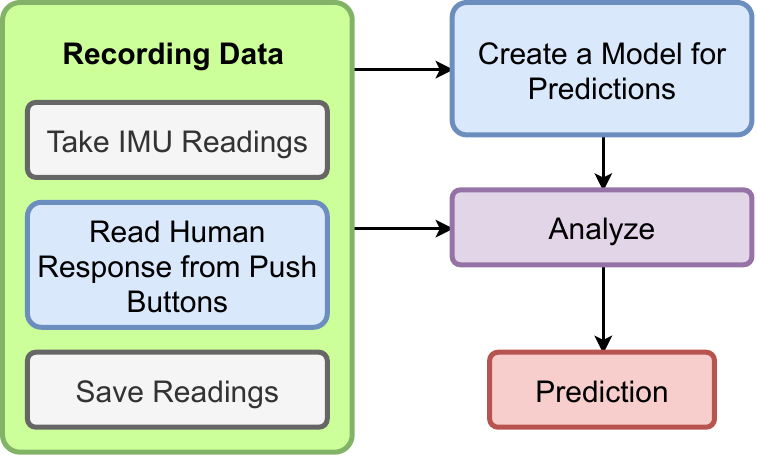}
\caption{Summarized description of the device}
\label{fig:flowchart}
\end{figure}

The device operates in two states, namely training state and prediction state: in the training state, it records both the sensor readings and the user inputs so that they can be used for creating the predictor models; in the prediction state, it reads sensor readings and uses the predictor models to make predictions, as depicted in~\figurename\ref{fig:flowchart}.

\section{Data Analysis}

\subsection{Preliminary readings}

\begin{figure}[!t]
\centering
\includegraphics[width=0.48\textwidth]{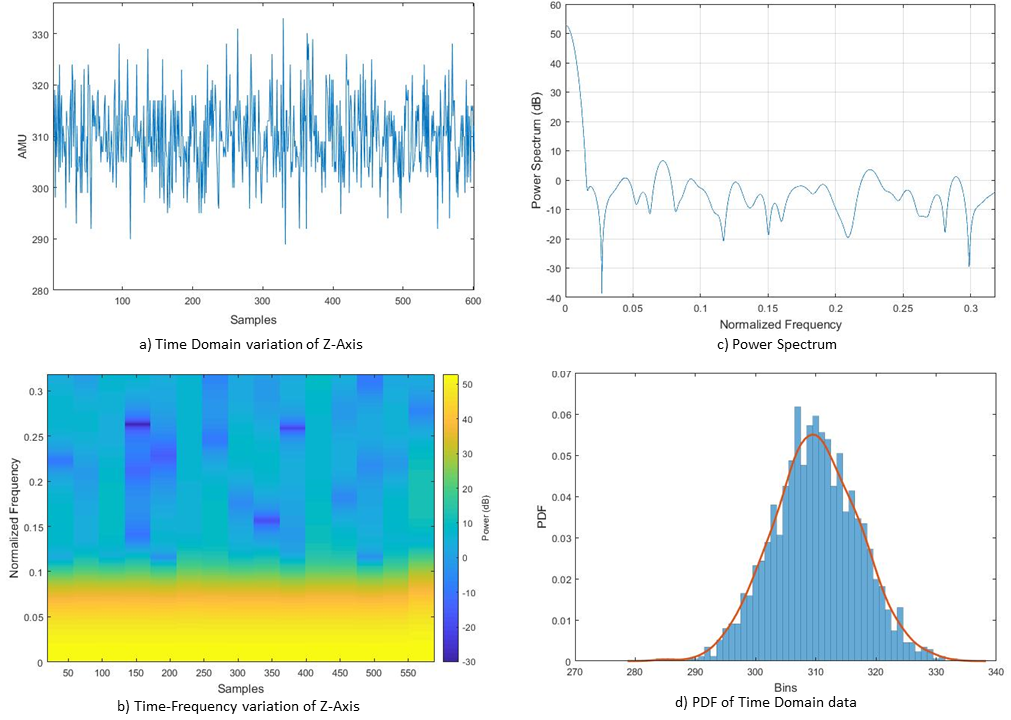}
\caption{Different representation of the data obtained while the device is at an stationery position}
\label{fig:rest}
\end{figure}

Following the fabrication of the system, several preliminary tests were conducted on the system to identify its performance in different environmental conditions as well as how it responds to different behaviours of pregnant mothers. Initially, the noise profile of the device is obtained. This is done by taking readings from the device while it's at a stationary position. The time-domain accelerometric readings of a single axis, as well as the time-frequency variation of the readings, can be observed in \figurename~\ref{fig:rest}a and \figurename~\ref{fig:rest}b. Moreover, the Probability Distribution Function (PDF) of the time domain data and the Power spectrum of the data can be observed in \figurename~\ref{fig:rest}c and \figurename~\ref{fig:rest}d, respectively. Furthermore, the kurtosis values of several samples were evaluated and the mean of the values were 2.9854, which is approximately 3.  Therefore, it can be observed that the stationery noise have a behaviour similar to Gaussian White noise.

\begin{figure}[!t]
\centering
\includegraphics[width=0.4\textwidth]{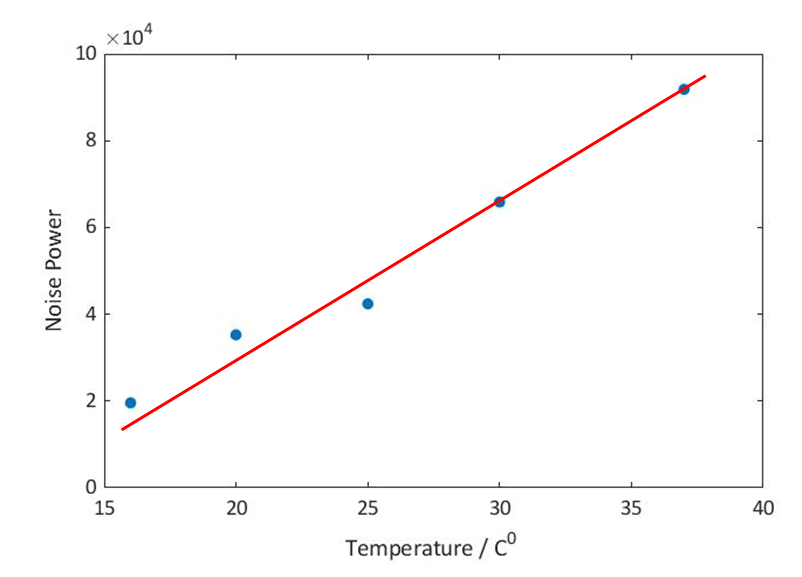}
\caption{Variation of noise power in different Temperature levels}
\label{fig:temp}
\end{figure}

The environmental temperature in the region where the studies were conducted varied from 16\degree C to 32\degree C. However, the usual body temperature is 37\degree C. Hence, when the sensor is worn for an extended time, the temperature could vary from 16\degree C to 37\degree C. In order to observe the effect of temperature on the noise features of the sensor, readings on different temperatures within the range of 16\degree C to 37\degree C were taken. The effect of the temperature can not be clearly observed in the time domain. Therefore, the noise power level at each temperature was calculated and plotted against the temperature to observe the effects. It can be observed in  \figurename~\ref{fig:temp} that when the temperature is increased, the noise power also increases gradually.

\begin{figure}[!t]
\centering
\includegraphics[width=0.48\textwidth]{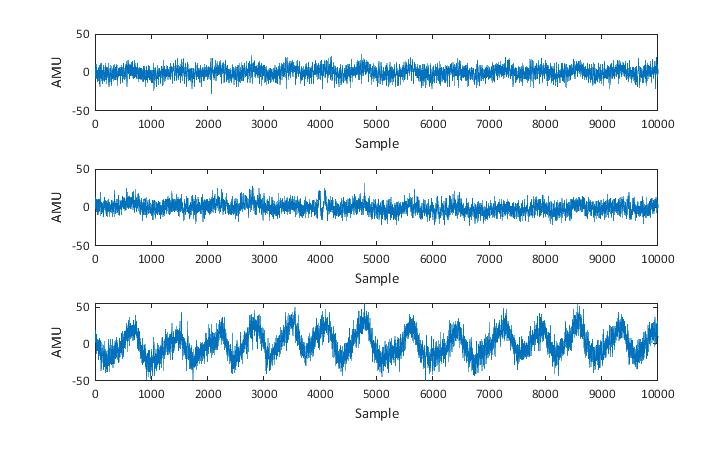}
\caption{When the device is worn by a stationery subject}
\label{fig:seated}
\end{figure}

Then the device was worn by a non-pregnant subject and readings were taken to observe the sensor response to different types of human behaviour. Initially, the subject was advised to wear the belt and to stay seated in the Fowler's position for an extended time and the tri-axial accelerometric data were recorded (see \figurename~\ref{fig:seated}). It can be observed that the subject's breathing pattern can also be observed very clearly from the accelerometric signals. While X-Axis and Y-Axis show a slight pattern, the Z-Axis indicate a clear pattern for the subject's breathing. Hence, it can be concluded that this body-worn accelerometer can also be utilized to monitor the subject's breathing patterns and further breathing anomalies. Furthermore, the effect of the position of the subject on the readings was observed by taking reading while the subject was in different positions such as the Fowler's position, Supine position, and lateral recumbent position. However, it was observed that the initial position of the subject do not have a significant effect on the data.

\begin{figure}[!t]
\centering
\includegraphics[width=0.48\textwidth]{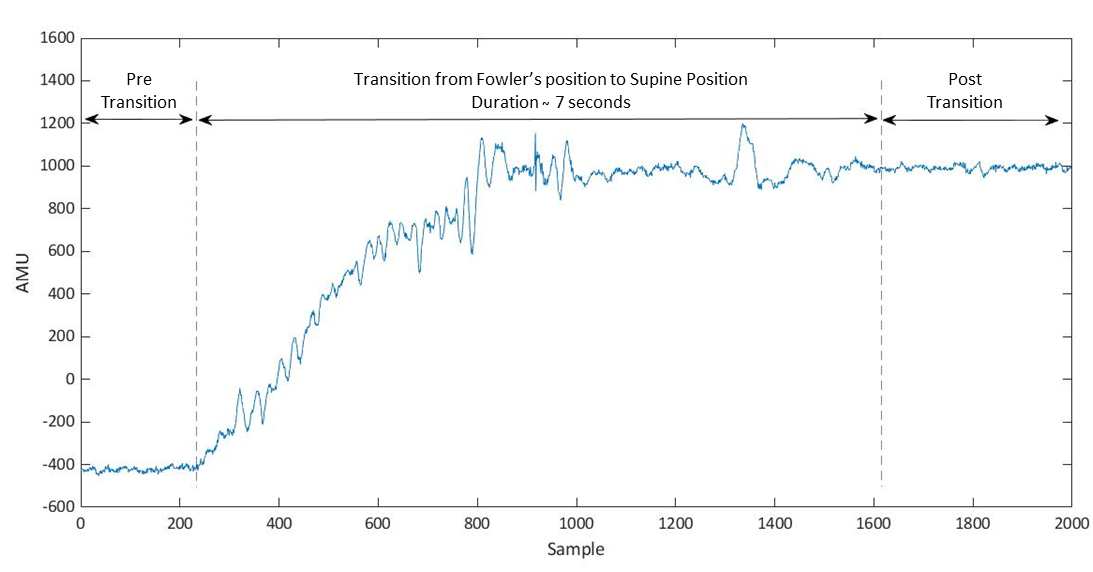}
\caption{Acceleration variation of the Z axis when the subject transit from Fowler's position to Supine position}
\label{fig:trans}
\end{figure}

However, since the main aim of this research is to conduct a preliminary study on the use of an accelerometric sensor-based system to monitor fetal and maternal health in the home, the effect of maternal movements on the sensor data was also observed. Initially, the effect on time-domain data when the subject was changing from one position to another position was observed. The subject was advised to change into different positions and the tri-axial accelerometric data were observed. Different transitions have different effects on each axis's readings. However, in all cases, it was observed that during the transition, there is a shift in the time domain data. It was also observed that following the transition period, the time domain data have a similar variation to time-domain data prior to transition. This can be observed in \figurename~\ref{fig:trans} where the Z-Axis variation when the subject is moving from Fowler's position to Supine position is depicted. Therefore, it can be concluded that such small movements (in time scale) of the subject will not have a significant effect on a session. The observed shift can be easily eliminated by utilizing simple signal processing techniques in the pre-processing stage. However, if a fetal movement is to coincide with a maternal movement, the extraction of the fetal movement signal may be strenuous. Nevertheless, due to the short duration of such activities, in practical situations, the probability of such coinciding occurrences can be considered to be very small.  

\begin{figure}[!t]
\centering
\includegraphics[width=0.48\textwidth]{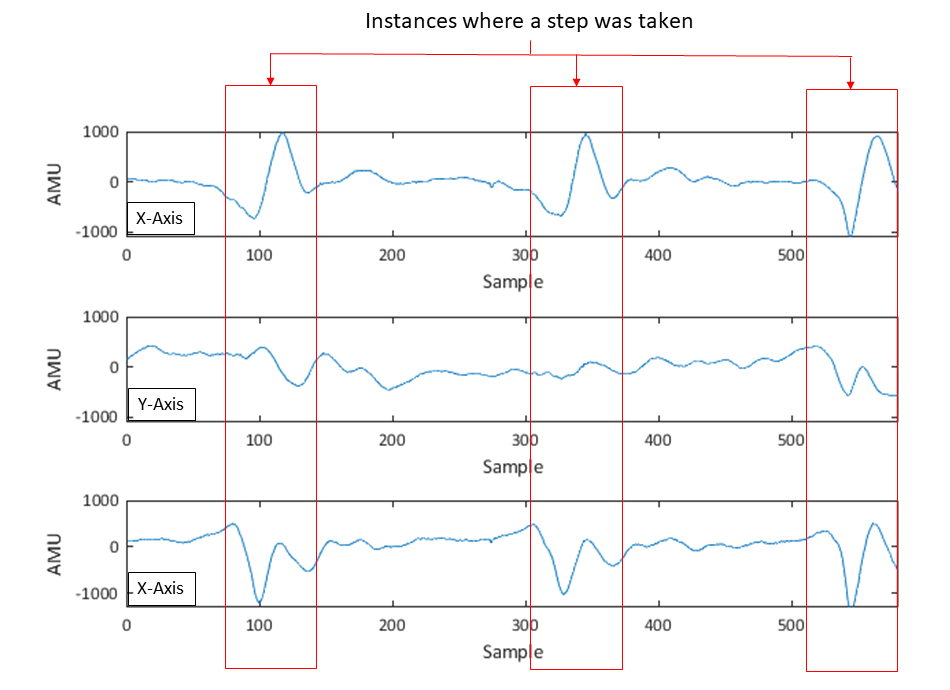}
\caption{Tri-Axial acceleration variation when the subject is walking while wearing the device}
\label{fig:steps}
\end{figure}

Thereafter, the effect of frequent maternal movements such as walking and talking were observed. The readings taken while the subject is walking can be observed in \figurename~\ref{fig:steps}. It can be observed that taking steps have a direct impact on the time domain data of all the axes. As discussed in the latter part of this paper, filtering methods such as Wiener filters can be applied to remove the imposed interference due to taking steps. However, it is advisably for mothers to stay stationary during a session. Furthermore, time-domain data obtained during the subject is speaking was compared with the data obtained while not speaking. No noticeable differences were observed. Therefore, it was concluded that speech has insignificant effect on the accelerometric data.

\begin{figure}[!t]
\centering
\includegraphics[width=0.48\textwidth]{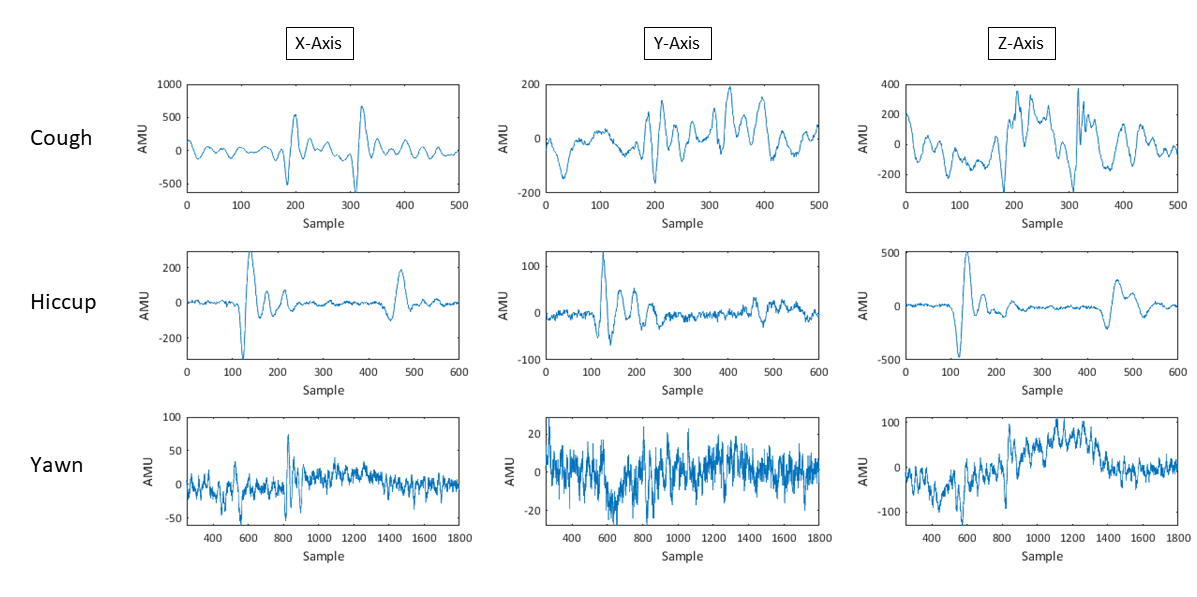}
\caption{Acceleration variation due to cough, Hiccup and yawn}
\label{fig:chy}
\end{figure}

Subsequently, the time domain accelerometric variation of several maternal motions such as cough, hiccup, yawn, and laugh were observed. These were selected due to the fact that they may have a similar effect on the maternal abdomen surface as of fetal movement. It can be observed in \figurename~\ref{fig:chy} that while cough, laugh, and hiccup have peaks similar to the fetal movement signal in the time domain, yawns do not have a similar time-domain profile. Hence, during analysis, more emphasis must be paid to extracting fetal movements signals from data contaminated with a maternal laugh, hiccups, and cough.  

\subsection{Respiratory data analysis}

\begin{figure}[!t]
\centering
\includegraphics[width=0.48\textwidth]{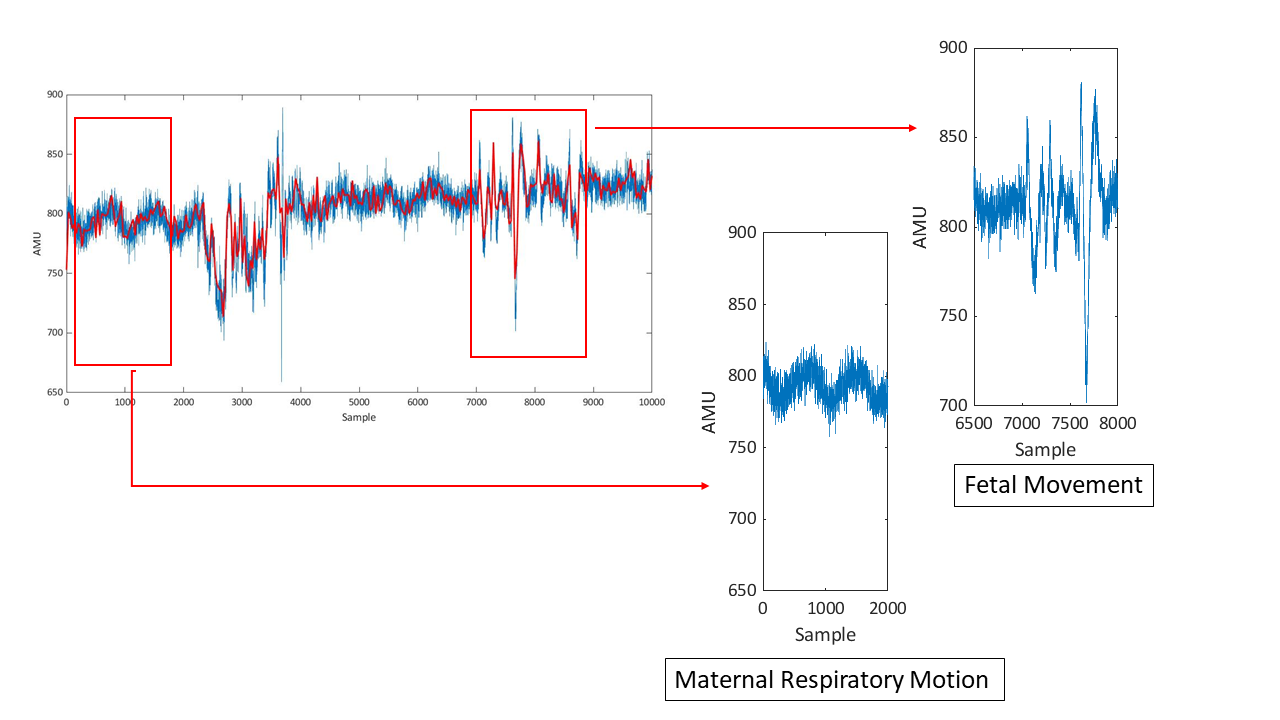}
\caption{Z-Axis Accelerometric variation of maternal respiratory motion and Fetal movement}
\label{fig:res}
\end{figure}

When observing the accelerometric data, it was noted that maternal respiratory movements are also present in the data. Furthermore, during the initial analysis of the device, the respiratory motion was observed clearly as can be seen in \figurename~\ref{fig:res}. In \figurename~\ref{fig:res} the time domain data of the Z-Axis is given and the general trend of the data is indicated in red colour.

Respiratory rate is one of the major well-being indicators of the human body. While the respiratory rate is usually overlooked, documenting respiratory rate can aid in predicting several serious clinical events \cite{res_imp}. Therefore, the viability of using this device to identify the respiratory rate was also studied.

In \figurename~\ref{fig:seated}, it can be observed that respiratory motion cause noticeable peaks in the Z-Axis data in the time domain. Therefore, it was decided to monitor these occurrences of peaks in the time domain in order to identify the respiratory rate. Initially, a de-trending algorithm was applied to the data to remove the mean as well as the trend in time domain data. In \figurename~\ref{fig:seated} and \figurename~\ref{fig:res}, it can be observed that the respiratory motion have a magnitude of approximately 50 AMU and the peaks are approximately 600 samples apart. This reading is taken while the subject was wearing the device at a stationary position. Further readings were taken after conducting two physical activities, a 10 minute walk and a 10 minute run. In all these readings, the magnitude of the respiratory motion was always less than 200 AMU and the peaks were more than 400 samples apart. Using this information, a peak identifying algorithm was developed and implemented and subsequently, the identified number of peaks were divided by the time they occurred to obtain the respiratory rate.

Furthermore, the assessment of the energy expenditure of the human body is considered to be of importance in sports as well as in other activities \cite{res_e2}. In \cite{res_e1}, it is discussed how to estimate human body energy expenditure by monitoring the respiratory patterns of the subject. They have utilized an accelerometric sensor and have conducted studies on three levels of energy expenditure; low EE, median EE, and high EE. In this study, it was studied how this sensor system can be utilized in identifying these three types of energy expenditure activities. 

Initially, three types of tasks were selected for the three levels of energy expenditure. Then an algorithm was implemented to identify the energy expenditure level of the body based on the respiratory motion data. A three-stage algorithm was utilized, where initially the data were prepossessed, then features were constructed, and finally, the extracted features were utilized in the classification of data. Initially, the tri-axial data were de-trended and the mean was removed. Then, (\ref{eqn_1}) was applied to the tri-axial data to eliminate sensor rotation interference and to combine data of the three axes \cite{acc_rot}. 

\begin{equation}
\label{eqn_1}
g(n) = \sqrt{x(n)^2 + y(n)^2 + z(n)^2}
\end{equation}

After prepossessing the data, the time domain signal was segmented into epochs of 1000 samples with a 20\% overlap. Then features were calculated for each epoch. When selecting the features to best represent the data, available time-domain features, as well as frequency domain features, were considered. More attention was paid to selecting a set of features that are not redundant as well as inclusive. It can be observed in \figurename~\ref{fig:res} that the time-domain data are very noisy. Therefore, it was decided to utilize frequency domain features.

When extracting features from the data in the frequency domain, initially, the discrete wavelet transform was applied to the time domain data, and it was segmented into four frequency bands \cite{dwt}. Individual features from each band were then computed. The features computed were; mean, standard deviation, variance, and the skewness of each band. From this, for a single epoch, 16 features were calculated and these features were fed into the classifying algorithm. However, in order to compare the performance of different types of wavelets on accelerometric signal classification, several types of wavelets (Daubechies wavelet db2, db4, db6, Symmlet wavelet sym2, and sym4) were used and the accuracy of each one was compared.

\begin{figure}[!t]
\centering
\includegraphics[width=0.48\textwidth]{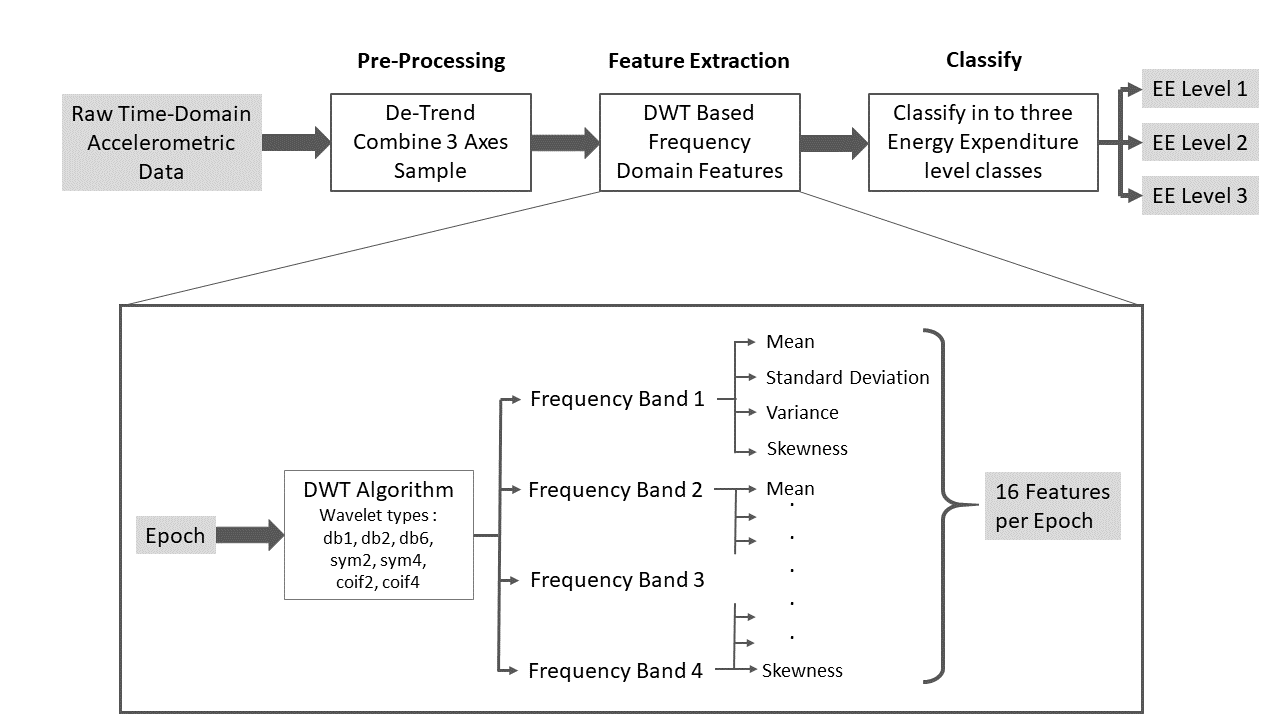}
\caption{Algorithm implemented to classify Energy Expenditure using respiratory data}
\label{fig:alg}
\end{figure}

The resulting features were fed into a simple standard neural pattern recognition algorithm. The input was set to be the features constructed in the previous step and the output was the three classes of activities. This network is a feed-forward network and the training was done utilizing the scaled conjugate gradient backpropagation. From the data set, 70\% of the data were used for training, 15\% were used for validation, and the remaining 15\% was used for testing. The summary of the implemented algorithm is illustrated in \figurename~\ref{fig:alg}.

\subsection{Fetal movement detection}

The data set obtained using the mothers was used for the analysis involving fetal movement detection. The analysis can be conducted in several different ways as it has been stated from rudimentary methods where the signal is observed for peaks to the use of machine learning algorithms. This paper focuses on using machine learning techniques since they can be implemented in small handheld devices with accurate results as it has been demonstrated in many wearables that are currently in wide use \cite{wearable}. 

The use of convolutional neural networks (CNNs) and recurrent neural networks (RNNs) are considered here. CNNs have proven to have generally good results for fetal movement identification as demonstrated in \cite{plos_one}, here spectrogram images obtained from the accelerometer readings were used as the input to the CNN. In this paper, to explore a different perspective, wavelet transform was used to generate the images. As stated in \cite{fftvswt}, wavelet transform suits better for non-stationary signal analysis when compared with Fourier analysis. The network used consisted of three convolution layers, each with 32 filters with a kernel size of 3, each followed by a max-pooling layer and finally a dense layer of 120 units.

Another approach that is considered here is the use of RNNs. RNNs have proven to perform best with the use of spectrograms in \cite{fetal_lstm} for fetal movement detection. Here, as an extension to that research, the spectrograms were used as the input to the network, but instead of using Long Short - Term Memory (LSTM) networks, Gated Recurrent Unit (GRU) networks were considered, which is almost similar in performance while being less computationally intensive \cite{gru_1}. The network implemented here consisted of 2 GRU layers along with dense, batch normalisation, and a 1-dimensional convolution layer.

\subsection{Interference removal}
The objective of this research is to develop a device to be used at home by mothers, without medical supervision. Hence, the posture and movements of the mother will not be as restricted as they would be in a hospital environment. This will require the device to have the ability to filter any interferences.

One of the common disturbances that can occur is due to walking, if the motion that the accelerometer picks up due to the physical movement is considered as a noise, this can be filtered to obtain the signal for breathing pattern analysis. Both breathing and walking patterns are of a periodic nature, and the Weiner filter would be a good choice to filter and obtain the pure breathing signal~\figurename~\ref{fig:weiner}. The accelerometer readings obtained while seated can be used as the reference signal.

\section{Results}

\subsection{Respiratory Data Analysis}

\begin{figure}[!t]
\centering
\includegraphics[width=0.48\textwidth]{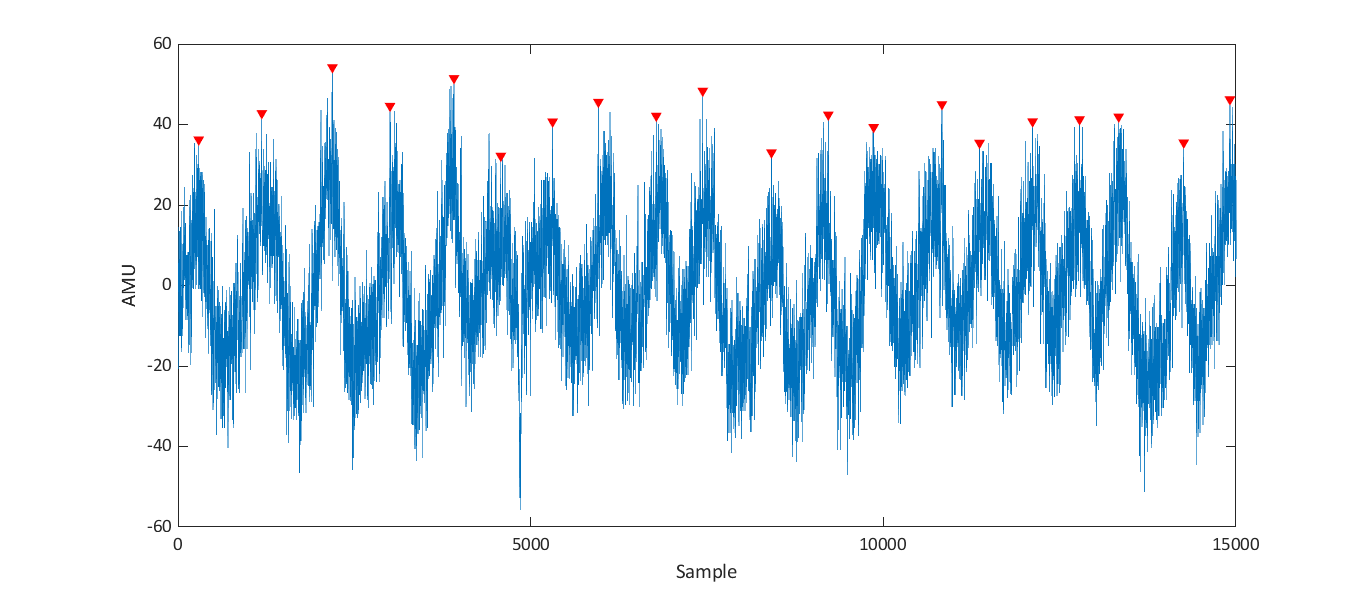}
\caption{Peaks detected in the Z-Axis Accelerometric data of respiratory motion}
\label{fig:peak}
\end{figure}

Initially an algorithm was implemented to identify the peaks of the repiratory motion signal. The identified peaks can be observed in \figurename~\ref{fig:peak}. The identified peaks are indicated using red arrowheads.

\begin{figure}[h]
\centering
\includegraphics[width=0.48\textwidth]{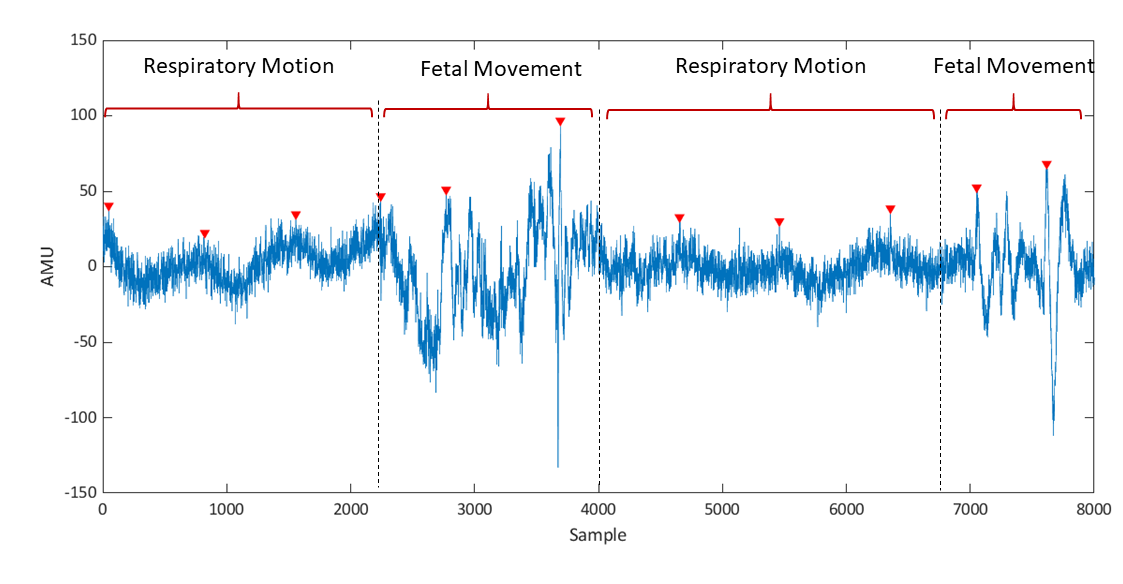}
\caption{Peaks detected when the signal contain fetal movement signal}
\label{fig:peak_fet}
\end{figure}

However, upon further investigation it was observed that since the target of the algorithm is to detect peaks, it may identify peaks at occurrence of a fetal movement as well. This can be observed in \figurename~\ref{fig:peak_fet}. When considering about the frequency of occurrence of fetal movement, the error introduced to the respiratory rate due to fetal movement can be considered to be negligible.

Subsequently, it was attempted to estimate the energy expenditure level utilizing respiratory data. During this, initially preprocessing and data segmentation was conducted. Thereafter, discrete Wavelet Transform was implemented on individual epochs to segment the time domain data into four frequency bands and four features were extracted from each band. These features were then used in the classification algorithm. When applying the DWT, different types of Wavelet signals were used and their performance was evaluated to identify the best wavelet to be utilized. Following the feature extraction, classification was conducted and the True Positive rate for each wavelet type is obtainded. The True positive rate obtained during training, validation, testing are given in Table \ref{tab:table1}.

\begin{table}[!t]
\caption{True positive rate obtained for different Wavelet types\label{tab:table1}}
\centering
\begin{tabular}{||c c c c c||} 
 \hline
 Wavelet & Training & Validation & Testing & Total \\ [0.5ex] 
 \hline\hline
 Db2 & 99.8 & 96.9 & 90.6 & 98.1 \\ 
 Db4 & 98.7 & 100 & 96.9 & 98.6 \\
 Db6 & 99.3 & 100 & 93.8 & 98.6 \\
 Sym2 & 96.7 & 93.8 & 90.6 & 95.4 \\
 Sym 4 & 99.2 & 98.9 & 93.8 & 98.9 \\ [1ex] 
 \hline
\end{tabular}
\end{table}


\subsection{Fetal movement detection}
\begin{figure}[!t]
\centering
\subfloat[]{\includegraphics[width=0.8\columnwidth]{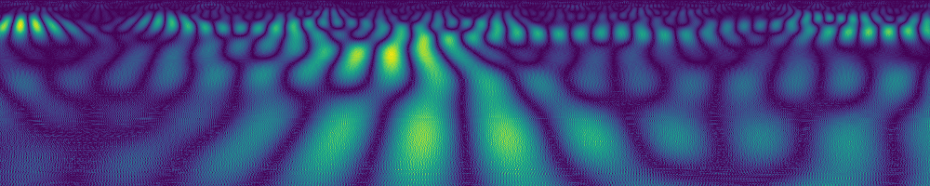}
\label{fig:wavelet_1}}
\hfil
\subfloat[]{\includegraphics[width=0.8\columnwidth]{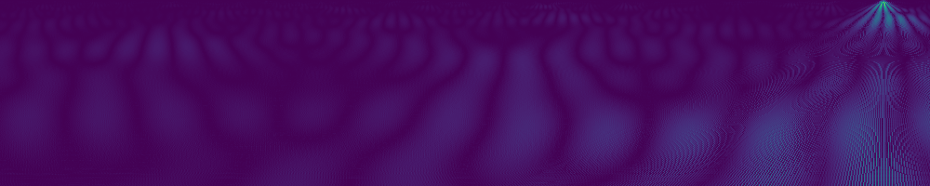}
\label{fig:wavelet_0}}
\caption{Wavelet scalograms generated from the data.(a) Scalogram with fetal movement. (b) Scalogram without fetal movement}
\label{fig:wavelet}%
\end{figure}

The wavelet transforms generated to be used as the inputs to the CNN showed significant visual differences for the two instances of the presence and absence of fetal movement~\figurename\ref{fig:wavelet}, although this was not as evident in some instances. 
Although the network trained showed good results for the original data set with higher accuracies (90\%), significant performance degradation can be observed if the validation of the results were done using a different mother. Hence, it can be inferred that the fetal movement patterns are somewhat unique to specific cases. Therefore, for better generalization, a significantly larger dataset must be considered or the model should be fine-tuned so that it fits each mother individually.

\begin{figure*}[!t]
\centering
\includegraphics[width=0.8\textwidth]{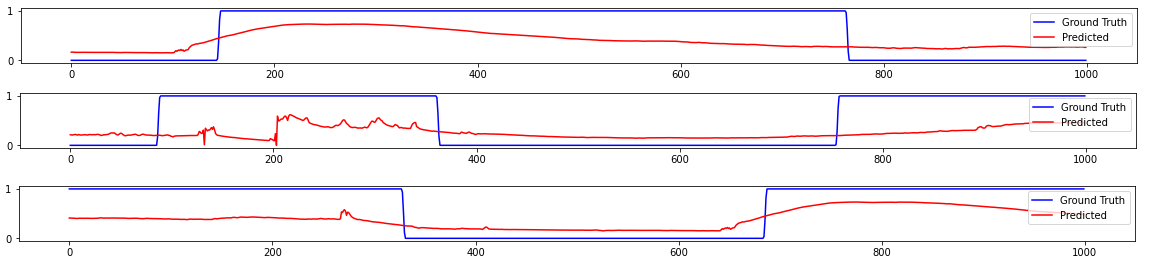}
\caption{Comparison between predicted and ground truth values.}
\label{fig:rnn_pred}
\end{figure*}

The same observation can be seen when an RNN was used, but with slightly better accuracies, which may be due to the fact that it gave more emphasis on the variation of the signal over time. A mean square error of 0.1 was observed by the final model. Also, the performance of the model compared to the ground truth is given in~\figurename\ref{fig:rnn_pred}.

\subsection{Interference removal}
\begin{figure}[!t]
\centering
\includegraphics[width=0.48\textwidth]{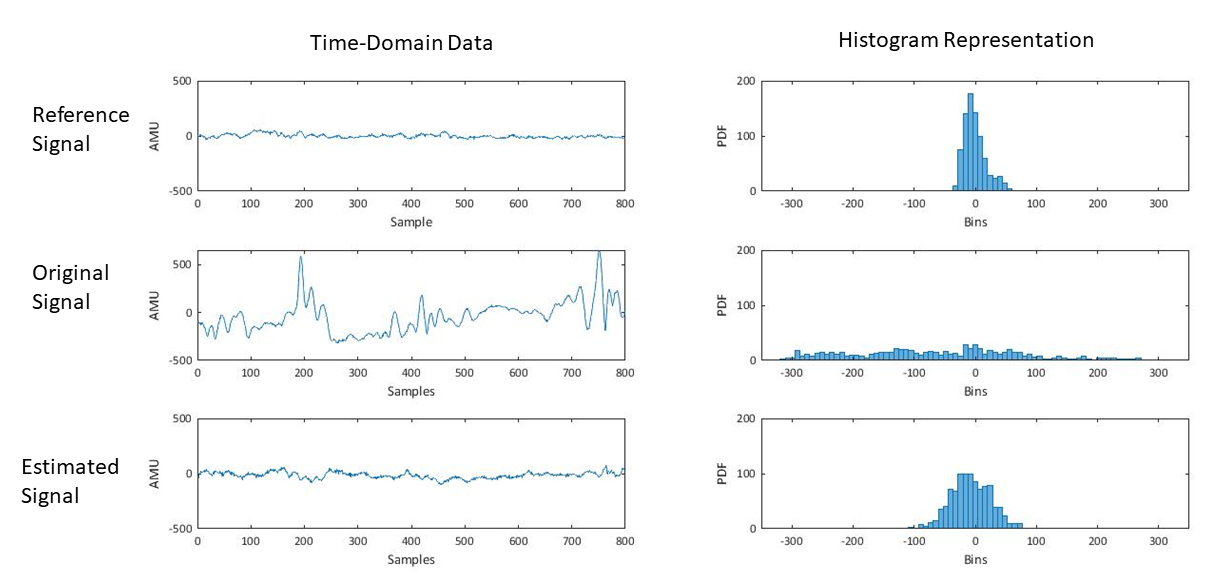}
\caption{Comparison between filtered and reference data}
\label{fig:weiner}
\end{figure}

Breathing patterns extracted from the accelerometer readings while walking did correspond to the accelerometer signals that were obtained while seated~\figurename\ref{fig:weiner}.

The algorithms, although not ideal, have been able to extract relevant data by filtering out the noise from walking, which has a much larger amplitude.

\section{Conclusion}
In conclusion, the simplicity of the device makes it cheap and easy to use and therefore the mother can have an active role in her health monitoring and can provide the medical experts with valuable information. This is mainly because the device depends on different filtering and analysis techniques rather than on complex hardware. Here, we have proposed the device to be used to monitor the respiratory patterns of the mother and fetal movement. Both of these metrics can provide valuable information when measured over a long period, which is not possible in the existing methods and the doctors have had to solely depend on the mother's observations which can be inconsistent and unreliable.

The device, along with the stated different analysis methods, can be implemented to obtain reasonably accurate results for pregnant mothers outside the hospital environment. It can give a rough estimate of the mother's and baby's health. Moreover, experts can observe the data from the device for further analysis as well. This device, therefore, allows a more personalised and long term monitoring solution that, although not highly accurate, can be a good addition to the existing health monitoring systems.\balance
\bibliographystyle{IEEEtran}
 \newcommand{\noop}[1]{}

\end{document}